\documentclass[11pt]{article}
\pdfoutput=1
\usepackage[centertags]{amsmath}
\usepackage[square, comma, sort&compress,numbers]{natbib}
\usepackage{array,multirow}
\numberwithin{equation}{section}
\usepackage{amssymb,amsfonts}
\usepackage{graphicx}
\usepackage{color}
\usepackage{mathtools,bm}
\usepackage{accents}
\usepackage{soul}

\usepackage{epsfig}
\usepackage{bbold}

\usepackage{wrapfig}
\usepackage{float}
\usepackage{soul}
\usepackage{tkz-euclide}
\usepackage{braket}

\usepackage{tikz,pgf}
\usetikzlibrary{shapes}
\usetikzlibrary{calc}
\usetikzlibrary{decorations.pathmorphing}
\usetikzlibrary{decorations.pathreplacing,shapes.misc}
\usetikzlibrary{positioning}
\usetikzlibrary{arrows}
\usetikzlibrary{decorations.markings}
\usetikzlibrary{shadings}

\usetikzlibrary{intersections}

\def\be{\begin{equation}}
\def\ee{\end{equation}}
\def\ba{\begin{array}}
\def\ea{\end{array}}

\def\bst{\begin{split}}
\def\est{\end{split}}

\def\dps{\displaystyle}

\def\1{\tilde{1}}
\def\2{\tilde{2}}
\def\3{\tilde{3}}

\newdimen\tableauside\tableauside=1.0ex
\newdimen\tableaurule\tableaurule=0.4pt
\newdimen\tableaustep
\def\phantomhrule#1{\hbox{\vbox to0pt{\hrule height\tableaurule
width#1\vss}}}
\def\phantomvrule#1{\vbox{\hbox to0pt{\vrule width\tableaurule
height#1\hss}}}
\def\sqr{\vbox{%
\phantomhrule\tableaustep

\hbox{\phantomvrule\tableaustep\kern\tableaustep\phantomvrule\tableaustep}%
\hbox{\vbox{\phantomhrule\tableauside}\kern-\tableaurule}}}
\def\squares#1{\hbox{\count0=#1\noindent\loop\sqr
\advance\count0 by-1 \ifnum\count0>0\repeat}}
\def\tableau#1{\vcenter{\offinterlineskip
\tableaustep=\tableauside\advance\tableaustep by-\tableaurule
\kern\normallineskip\hbox
{\kern\normallineskip\vbox
{\gettableau#1 0 }%
\kern\normallineskip\kern\tableaurule}%
\kern\normallineskip\kern\tableaurule}}
\def\gettableau#1 {\ifnum#1=0\let\next=\null\else
\squares{#1}\let\next=\gettableau\fi\next}

\newcommand{\bref}[1]{\textbf{\ref{#1}}}

\newcommand{\re}{\mathop{\mathrm{Re}}}

\def\cD{\mathcal{D}}

\def\cF{\mathcal{F}}

\def\cH{\mathcal{H}}

\def\cJ{\mathcal{J}}

\def\cL{\mathcal{L}}

\def\cO{\mathcal{O}}

\def\cV{\mathcal{V}}

\numberwithin{equation}{section} \makeatletter
\@addtoreset{equation}{section}

\def\be{\begin{equation}}
\def\ee{\end{equation}}
\def\ba{\begin{array}}
\def\ea{\end{array}}

\def\dps{\displaystyle}

\def\ba{\begin{array}}
\def\ea{\end{array}}

\def\dps{\displaystyle}

\def\cft{CFT$_D$ }

\def\hd{\frac{D}{2}}

\newcommand{\dl}{\Delta}
\newcommand{\td}{\tilde{\Delta}}

\def\Li2{\operatorname{Li_2}}

\def\Tr{\operatorname{Tr}}
\def\C2{\text{C}_2}

\def\cylR{\mathbb{R}\times \mathbb{S}^{D-1}} 
\def\cylS{\mathbb{S}_\beta^1\times \mathbb{S}^{D-1}}

\usepackage{jheppub}
\makeatletter
\def\@fpheader{\vspace{-.1cm}}
\makeatother

\title{\centering{One-point thermal conformal blocks \\from four-point conformal integrals}}

\author{Konstantin\ Alkalaev}   
\author{and Semyon\ Mandrygin}

\affiliation{I.E. Tamm Department of Theoretical Physics, \\P.N. Lebedev Physical
Institute, 119991 Moscow, Russia}

\emailAdd{alkalaev@lpi.ru}
\emailAdd{semyon.mandrygin@gmail.com}

\abstract{We develop the thermal shadow formalism to study  the conformal blocks decomposition in $D$-dimensional conformal field theory on $\mathbb{S}_{\beta}^{1} \times \mathbb{S}^{D-1}$, where the temperature is $T = \beta^{-1}$. It is demonstrated that both the 1-point thermal ($T\neq 0$) conformal blocks and the 4-point plane ($T=0$) conformal blocks are defined by the same 4-point conformal integral. It is shown that up to power prefactors the 1-point thermal conformal block is given by the fourth Appell function.}

\begin{document}

\maketitle
\flushbottom

\section{Introduction}

Conformal field theory at finite temperature $T=1/\beta$ has been of interest since  the earliest  stages of  development of the holographic correspondence \cite{Witten:1998zw}.  At zero temperature, significant progress in solving \cft has been achieved  by using the bootstrap prorgam, for a review see e.g. \cite{Poland:2018epd, Bissi:2022mrs}. Within this approach, the conformal blocks are instrumental   and several powerful methods have been developed to calculate them effectively, see e.g. \cite{Dolan:2011dv, Costa:2011dw,Karateev:2017jgd,Rosenhaus:2018zqn,Fortin:2019zkm, Poland:2021xjs,Buric:2021ywo, Ammon:2024axd}. However, it is not the case with \cft at non-zero temperature, where only partial results on thermal conformal blocks are available \cite{Iliesiu:2018fao, Gobeil:2018fzy, Manenti:2019wxs}.

Being aimed at finding efficient methods of calculating thermal blocks,  in this paper we adapt the  shadow formalism \cite{ Ferrara:1972ay, Ferrara:1972xe, Ferrara:1972uq, Ferrara:1972kab} for  thermal CFT$_D$. In this regard, our constructions can be viewed as a natural generalization of the torus CFT$_2$ shadow formalism \cite{Alkalaev:2023evp} (for higher-spin torus CFT$_2$ generalization see \cite{Belavin:2024nnw}). Note that in $D$ dimensions  one is usually interested in considering  a thermal theory on one of two backgrounds, $S^1_{\beta}\times \mathbb{R}^{D-1}$ or $\mathbb{S}_{\beta}^{1} \times \mathbb{S}_{L}^{D-1}$, where $\beta$ and $L$ are radii. These two spaces are related by (de)compactification of the second factor: e.g. sending  $L\to \infty$ in $\mathbb{S}_{\beta}^{1} \times \mathbb{S}_{L}^{D-1}$ we obtain  $\mathbb{S}^1_{\beta}\times \mathbb{R}^{D-1}$. From this perspective, a thermal \cft on $\mathbb{S}_{\beta}^{1} \times \mathbb{S}_{L}^{D-1}$ is more general.\footnote{Due to (residual) conformal invariance all observables, including correlation functions, depend only on the ratio $\beta/L$. For brevity, we set the radius $L=1$ (which can be recovered at any stage by dimensional argument).}

Using the thermal shadow formalism we explicitly calculate the 1-point conformal block for scalar operators on $\cylS$ and find out that the resulting function can be naturally represented as the forth Appell function with dependent arguments. In general, our basic observation is that the thermal conformal blocks are defined by a particular limit of conformal integrals which in their turn also define conformal blocks on $\mathbb{R}^{D}$. In this way, one concludes that knowing the explicit form of multipoint conformal integrals in $D$ dimensions  solves the problem of finding conformal blocks at both  zero and non-zero temperatures.

The paper is organized as follows. In section \bref{sec:CFT} 1-point correlation functions, thermal conformal blocks, (broken) conformal symmetries, and the corresponding Ward identities in thermal \cft are briefly  discussed. In section \bref{sec:shadow} we develop the thermal shadow formalism. Here, we explicitly calculate  scalar 1-point  thermal conformal block and study the resulting function.  In the concluding section \bref{sec:concl} we summarize our findings and discuss further tasks in thermal CFT$_D$. {Appendix \bref{app} considers the non-parametric conformal integral and the corresponding 1-point thermal block.}

\section{One-point correlation function at finite temperature}
\label{sec:CFT}

We consider a theory  at finite temperature $T=\beta^{-1}$  living on a cylinder $\cylR$ with the standard metric ${\rm d} s^2_{\cylR} = {\rm d}\tau^2+{\rm d} \Omega^2_{D-1}$ and define the 1-point correlation function of a scalar primary operator $\phi_{\cylR}(\tau, \Omega)$ of conformal dimension $h$ as
\be
\label{corrcyl}
\langle \phi_{\cylR} (\tau,\Omega) \rangle_\beta = \Tr_{\cH}\left[ \phi_{\cylR} (\tau,\Omega) e^{-\beta D} \right]\\.
\ee
Here, $D$ is a dilatation operator and the trace $\Tr$ is taken over the space of states\footnote{Within  the radial quantization the space of  states $\cH$ is  given by functions on $\mathbb{S}^{D-1}$ and the dilatation  $D$ acts as the Hamiltonian evolving (in time on a cylinder $\tau$) the states from one sphere to another.} $\cH = \oplus V_{\dl, {\bm s}}$, where $V_{\dl, {\bm s}}$ is a (generalized) Verma module for the conformal algebra $o(D+1,1)$ with conformal dimension $\Delta$ and generalized spin ${\bm s} = (s_1,..., s_{[\frac{D+2}{2}]})$. We will be considering only scalar Verma modules in which case ${\bm s}=0$;  then, $V_{\dl, {\bm s}} \equiv  V_{\dl}$ is a highest weight representation generated from a primary state $\ket{\dl}$ defined by 
\be
D \ket{\dl} = \dl \ket{\dl}\,,
\qquad
J_{\mu\nu} \ket{\dl} = 0\,,
\qquad
K_\mu \ket{\dl} = 0\,,
\ee
where the conformal algebra generators obey the commutation relations
\be
\ba{l}
{\left[D, P_\mu\right] } =P_\mu\,, \quad \left[D, K_\mu\right]=-K_\mu\,, 
\quad\left[K_\mu, P_\nu\right]=2 \delta_{\mu \nu} D-2 J_{\mu \nu}\;, 
\vspace{2mm} 
\\
\dps
{\left[J_{\mu \nu}, P_\rho\right] }  =\delta_{\nu \rho} P_\mu-\delta_{\mu \rho} P_\nu\,, \quad\left[J_{\mu \nu}, K_\rho\right]=\delta_{\nu \rho} K_\mu-\delta_{\mu \rho} K_\nu\, 
\vspace{2mm} 
\\
\dps
{\left[J_{\mu \nu}, J_{\rho \sigma}\right] }  =\delta_{\nu \rho} J_{\mu \sigma}-\delta_{\mu \rho} J_{\nu \sigma}+\delta_{\nu \sigma} J_{\rho \mu}-\delta_{\mu \sigma} J_{\rho \nu}\,,
\ea
\ee
with  $\delta_{\mu\nu}$ being  the $o(D)$ invariant tensor, $\mu, \nu, ... = 1,...,D$. The inner product is defined by the conjugation rules:  $D^{\dagger}=D$, $J_{\mu \nu}^{\dagger}=-J_{\mu \nu}$, $P_\mu^{\dagger}=K_\mu$, and $K_\mu^{\dagger}=P_\mu$. Descendant states in $V_{\dl}$:
\be
\label{descendants}
\ket{\dl+n}_{\mu_1 ... \mu_n} = P_{\mu_1} \ldots P_{\mu_n}\ket{\dl}\,,
\quad
D \ket{\dl+n}_{\mu_1 ... \mu_n} = (\dl+n)\ket{\dl+n}_{\mu_1 ... \mu_n}\,, 
\ee 
$n = 0,1,2,...$ is a level. Each level is reducible with respect to $o(D)$ subalgebra and decomposes into $[n/2]+1$ finite-dimensional irreducible $o(D)$ representations of spins $j  = n, n-2, n-4, ...$ with multiplicity 1.   

The thermal correlation function \eqref{corrcyl} is periodic in time coordinate $\tau$:  
\be
\label{kms}
\ba{l}
\langle \phi_{\cylR}(\tau,\Omega) \rangle_\beta = \Tr_{\cH}\Big[e^{\tau D} \phi_{\cylR}(0,\Omega) e^{-\tau D} e^{-\beta D} \Big]
\vspace{2mm}
\\
\dps
\hspace{20mm}= \Tr_{\cH} \Big[ e^{-\beta D} e^{(\beta+\tau) D} \phi_{\cylR}(0,\Omega) e^{-(\beta+\tau) D} \Big]  = \langle \phi_{\cylR}(\tau+\beta,\Omega) \rangle_\beta \,,
\ea 
\ee
where in the second line we inserted $\mathbb{1} = e^{-\beta D} e^{\beta D}$ and  used the cyclic property. It means that this correlation function  is actually  defined  on $\cylS$. When considering higher-point correlation functions the argument \eqref{kms} turns into KMS condition, serving as a crossing symmetry equation in the thermal bootstrap \cite{El-Showk:2011yvt,Iliesiu:2018fao}.  In fact, the inverse temperature $\beta$ is always present in thermal \cft as the geometric parameter. 

On the other hand, $\cylR$ is related to $\mathbb{R}^D$ with coordinates $x^{\mu}$ through the standard Weyl map $r= e^{\tau}$, where $r^2 = x^\mu x_\mu$. The transformation rule for a primary operator $\phi_{\cylR} (\tau,\Omega)= r^h \phi_{\mathbb{R}^D}(x)$ allows expressing  \eqref{corrcyl} as
\be
\label{corrconnect}
\langle \phi_{\cylR} (\tau,\Omega) \rangle_\beta = r^h \,  \Tr_{\cH}\left[ \phi_{\mathbb{R}^D}(x) e^{-\beta D} \right] \equiv r^h \langle \phi_{\mathbb{R}^D}(x)  \rangle_\beta \,.
\ee

In what follows we will study thermal correlation functions of scalar primary operators $\phi(x) \equiv \phi_{\mathbb{R}^D}(x)$ on $\mathbb{R}^D$ defined by \eqref{corrconnect}. Focusing on the contribution of scalar operators one can expand the thermal correlator $\langle \phi(x) \rangle_{\beta}$ in conformal blocks
\be
\label{blockexp}
\begin{split}
\langle \phi (x) \rangle_{\beta} \simeq \sum_{\dl} C_{\dl,h,\dl} \cF_{\dl}^{h}(q,x)\,,
\end{split}
\ee
the symbol $\simeq$ means that we omitted contributions of spinning operators. Each scalar thermal conformal block here is the series  in $q = \exp (-\beta)$
\be
\label{blockdef}
 \cF_{\dl}^{h}(q,x) = (C_{\dl,h,\dl})^{-1}\sum_{n=0}^{\infty} q^{\dl+n} \left(B^{-1}_{\dl}\right)^{\mu_1 \ldots \mu_n ; \nu_1 \ldots \nu_n} {}_{\nu_1 \ldots \nu_n }\langle \dl+n | \phi(x) |\dl+n  \rangle_{\mu_1\ldots \mu_n} \,,
\ee
where  $B_{\nu_1 \ldots \nu_n; \mu_1 \ldots \mu_n } = {}_{\nu_1 \ldots \nu_n }\langle \dl+n| \dl+n \rangle_{\mu_1\ldots \mu_n}$ is a Gram matrix in $V_{\dl}$ at $n$-th level. From \eqref{corrconnect} it follows that the scalar thermal conformal blocks in different coordinates are related as  $\cF_{\dl}^h(q,\tau,\Omega) = r^h \, \cF_{\dl}^{h}(q,x)$.

The 1-point thermal correlation function \eqref{corrcyl} can be generalized by adding chemical potentials
\be
\label{gen_cor}
\langle \phi(x) \rangle_{{\bm \beta}} \equiv \Tr_{\cH}\left[q_1^{H_1}\cdots  q_r^{H_r} \\ \phi(x) \right]  \,,
\ee
where $r = rank \,o(D+1,1) = [\frac{D+2}{2}]$ and $H_i$ are basis Cartan elements, ${\bm q} = (q_1, \ldots, q_r)$, where $q_i = \exp(-\beta_i)$ and ${\bm \beta} = (\beta_1, \ldots, \beta_r)$. Conformal invariance of \eqref{gen_cor} can be formulated as an invariance of the correlation function \eqref{gen_cor} under the action of a conformal transformation (at a given point $x$)
\be
\label{inv_cond}
\langle \phi(x) \rangle_{{\bm \beta}} = \langle U^{-1} \phi(x) U \rangle_{{\bm \beta}} \,,
\ee    
where $U$ is a linear operator representing a particular conformal group transformation. The generalized chemical potentials in \eqref{gen_cor} break $O(D+1,1)$ down to a subgroup. Indeed, inserting a number of resolutions of identities  $\mathbb{1} = U^{-1} U $ into \eqref{gen_cor} and using the cyclic property one finds that those $U$ which define a residual conformal symmetry satisfy the condition
\be
[U, H] = 0 \,,
\ee  
where $H \equiv q_1^{H_1}\cdots  q_r^{H_r} = \exp(-\beta_1 H_1) \cdots  \exp(-\beta_r H_r) $ denotes a particular element of the  Cartan subgroup of $O(D+1,1)$. In other words, admissible $U$ span a centralizer of the Cartan subgroup, $U\in C_{O(D+1,1)}(H)$.\footnote{Note that the same logic applies to the case of choosing ${\bm \beta} = {\bm 0}$ and boiling the trace down to sandwiching between  the vacuum states $\bra{0}$ and $\ket{0}$. The cyclic property of the trace is now replaced by the vacuum invariance condition $U\ket{0}=0$. Then, $U\in C_{O(D+1,1)}(\mathbb{1}) = O(D+1,1)$ and we recover the full conformal invariance of \cft on $\mathbb{R}^D$.} In our case ${\bm \beta} = (\beta,0,...,0)$ so that the centralizer is identified with  the subgroup  $O(1,1)\oplus O(D)$, which generators are the dilatation $D$ and the Lorentz rotations $J_{\mu\nu}$. Then, the infinitesimal version of \eqref{inv_cond} is the Ward identities:
\be
\label{ward}
\cD\,\langle \phi(x) \rangle_{{ \beta}} = 0\;,
\qquad
\cJ_{\mu\nu}\,\langle \phi(x) \rangle_{{ \beta}}  = 0\;,
\ee 
where $\cD$ and $\cJ_{\mu\nu}$ are differential operators which define the variations of $\phi(x)$ at a given point under the action of symmetry generators $D$ and $J_{\mu\nu}$. The Ward identities completely fix the coordinate dependence of the 1-point correlation function \eqref{corrcyl} (see the conformal block expansion \eqref{blockexp} and the final expresion \eqref{tblock}).\footnote{In the similar way, the 1-point correlation functions in torus CFT$_2$ depend on the toric modulus (temperature analog)  and not on local coordinates on the torus $\mathbb{S^1}\times \mathbb{S^1}$.} In particular, the first constraint  in \eqref{ward} claims that  it  does not depend on time $\tau$ (in cylindrical coordinates). Adding more chemical potentials  reduces the number of the Ward identities that, in particular, complicates $x$-dependence (see e.g. \cite{Gobeil:2018fzy}).\footnote{For a more detailed discussion of conformal symmetries at finite temperature and the respective Ward identities see e.g. \cite{Marchetto:2023fcw}.} Nonetheless, the 1-point correlation function $\langle \phi(x) \rangle_{{\bm \beta}}$ is still $\tau$-independent because $D$ is always among the residual symmetry generators.

{It is remarkable that the high-temperature limit $\beta \to 0$ of the correlation functions on $\cylS$ defines the correlation functions on the other thermal background $S^1_{\beta} \times \mathbb{R}^{D-1}$. Particularly, for a primary scalar operator $\phi(\tau, \vec{x})$, where  $\vec{x}\in\mathbb{R}^{D-1}$, the 1-point thermal correlation function  can be written as \cite{El-Showk:2011yvt,Iliesiu:2018fao} 
\be
\label{bh}
\langle \phi(\tau, \vec{x}) \rangle_{S^1_{\beta} \times \mathbb{R}^{D-1}} = \frac{b_h}{\beta^h}\,, \qquad b_h \,\simeq\, \lim_{\beta \to 0} \beta^h\, \frac{\sum_{\dl} C_{\dl,h,\dl} \cF_{\dl}^{h}(q,x)}{Z(\beta)}\bigg|_{r=1}\,,
\ee
where $Z(\beta) = \Tr e^{-\beta D}$ is a partition function on $\cylS$ (contributions of spinning operators are omitted). As will be discussed in section \bref{sec:1pt_thermal}, the thermal conformal blocks are drastically  simplified at $\beta \to 0$ and the (model-dependent) constant $b_h$, which defines thermal observables on $S^1_{\beta} \times \mathbb{R}^{D-1}$ is determined in terms of the plane \cft structure constants $C_{\dl,h,\dl}$.  For illustration of calculation of  $b_{h}$ in particular models (with chemical potentials and other theory parameters), see e.g. \cite{ Iliesiu:2018fao,Petkou:2021zhg}. }

\section{Thermal shadow formalism}
\label{sec:shadow}

For a scalar primary operator $\cO_{\dl}(x)\equiv \cO(x)$  one can define the shadow operator \cite{Ferrara:1972ay, Ferrara:1972xe, Ferrara:1972uq, Ferrara:1972kab, Fradkin:1978pp, Fradkin:1996is, SimmonsDuffin:2012uy} 
\be
\label{shadop}
\widetilde{\cO}(x) =N_{\dl}
\int_{\mathbb{R}^D} {\rm d}^D x_0 \, (x_0-x)^{-2\td}\, \cO(x_0)\,, \qquad N_{\dl}  =  \pi^{-D} \frac{\Gamma(\dl) \Gamma(\td)}{\Gamma(\hd-\dl) \Gamma(\hd-\td)} \,,
\ee
which is a primary operator of (dual/shadow) conformal dimension $\tilde{\dl} = D - \dl$; the power-law function  $(x_0-x)^{-2\td}$ is its propagator. With the chosen normalization factor $N_{\dl}$ the 2-point correlation function reads as  
\be
\label{shad2pt}
\langle \widetilde{\cO}(x_1) \cO(x_2) \rangle = \delta (x_1-x_2)\,,
\ee
so one can define the conformally invariant projecting operator which satisfy  the idempotent and orthogonality properties
\be
\label{shadproj}
\Pi_{\dl} = \int_{\mathbb{R}^D} {\rm d}^D x \, \cO(x) |0\rangle \langle 0| \widetilde{\cO}(x)\,,
\qquad
\Pi_{\dl_n} \Pi_{\dl_m} = \delta_{\Delta_n\Delta_m} \Pi_{\dl_m}\,.
\ee

Since $\widetilde{\cO}$ is a primary operator one finds the following 3-point function 
\be
\label{shad3pt}
\langle \phi_1(x_1) \phi_2 (x_2) \widetilde{\cO}(x_3) \rangle = C_{h_1,h_2,\td} V_{h_1,h_2,\td} (x_1,x_2,x_3) \,,
\ee
where $\phi_{h_i} (x_i) \equiv \phi_i(x_i)$ are primary operators of conformal dimensions $h_i$, $C_{h_i,h_j,h_k}$ are structure constants, and\footnote{It is convenient to use $\Delta$ and/or $\td$ for internal (shadow) dimensions and $h_i$ for external dimensions.} 
\be
\label{vert3}
V_{h_i,h_j,h_k}(x_i,x_j,x_k) = X_{ij}^{\frac{h_k-h_i-h_j}{2}} X_{ik}^{\frac{h_j-h_i-h_k}{2}} X_{jk}^{\frac{h_i-h_j-h_k}{2}}\,,
\quad\text{where}\quad
X_{ij} = |x_i-x_j|^2 \,.
\ee
On the other hand, using the integral formula \eqref{shadop} one can show that 
\be
\label{shad3ptstar}
\langle \phi_1(x_1) \phi_2 (x_2) \widetilde{\cO}(x_3) \rangle = C_{h_1,h_2,\dl} N_{\dl} X_{12}^{\frac{\dl-h_1-h_2}{2}} \,  \int_{\mathbb{R}^D} {\rm d}^D x_0 \, X_{01}^{\frac{-\dl-h_{12}}{2}}  X_{02}^{\frac{-\dl+h_{12}}{2}} X_{03}^{-\td} \,,
\ee
where $h_{ij}=h_i - h_j$. Recalling that $\tilde{\dl} = D - \dl$ one observes that the sum of powers of propagators under the integral equals the space dimension $D$. Thus, the integral is determined by the star-triangle relation \cite{Symanzik:1972wj} 
\be
\label{startriangle}
\int_{\mathbb{R}^D} {\rm d}^D x_0 \, X_{01}^{-a_1}  X_{02}^{-a_2} X_{03}^{-a_3} = \pi^{\hd} \Gamma 
\left[
\begin{array}{l l}
a'_1, a'_2, a'_3 \\
a_1, a_2, a_3
\end{array}
\right] 
X_{12}^{-a'_3} X_{13}^{-a'_2} X_{23}^{-a'_1}\,,
\ee
where $a_1+a_2+a_3 = D$ and $a' = \hd - a$,  the notation for $\Gamma$-functions is introduced
\be
\label{gammas}
\Gamma
\left[
\begin{array}{l l}
a_1, \ldots, a_n \\
b_1, \ldots, b_m
\end{array}
\right] = \frac{\Gamma(a_1, \ldots, a_n)}{\Gamma(b_1, \ldots, b_m)}\,, \qquad \Gamma(a_1, \ldots, a_n) = \prod_{i=1}^{n} \Gamma(a_i)\,.
\ee
For later purposes, comparing \eqref{shad3pt} and \eqref{shad3ptstar} we note that the structure constants $C_{h_1,h_2,\td}$ and $C_{h_1,h_2,\dl}$ are related as\footnote{The coefficient $K_{\dl}^{h_1,h_2}$ arises when decomposing the 4-point conformal partial wave into conformal and shadow blocks, see e.g. \cite{Rosenhaus:2018zqn}.}
\be
\label{structconstratio} 
\frac{C_{h_1,h_2,\td}}{C_{h_1,h_2,\dl}} = N_{\dl} K_{\dl}^{h_1,h_2} \,,
\quad
\text{where}
\quad
K_{\td}^{h_1,h_2} = \pi^{\hd} \, \frac{ \Gamma(\hd - \dl) \Gamma(\frac{\dl-h_{12}}{2}) \Gamma(\frac{\dl+h_{12}}{2}) }{\Gamma(\dl) \Gamma(\frac{\tilde{\dl}-h_{12}}{2}) \Gamma(\frac{\tilde{\dl}+h_{12}}{2})}\,,
\ee
and the product $ K_{\dl}^{h_i,h_j} K_{\td}^{h_i,h_j} = 1/N_{\dl}$ is independent of $h_i$ and $h_j$.

\subsection{One-point thermal conformal partial wave}

Let us see how to use  shadow operators in order to calculate the scalar conformal blocks in thermal CFT$_D$. To this end, one  inserts the projecting operator $\Pi_{\dl}$ \eqref{shadproj} into the 1-point thermal correlation function  \eqref{blockexp}-\eqref{blockdef}, considering only the scalar operator contributions
\be
\label{trprojfirst}
\begin{split}
\Tr_{\cH} \left[\Pi_{\dl} \phi(x) q^{D} \right] &= \int_{\mathbb{R}^D} {\rm d}^D x_0 \,\sum_{\dl_1} \,\sum_{n=0}^{\infty}  \left(B^{-1}_{\dl_1}\right)^{\mu_1 \ldots \mu_n ; \nu_1 \ldots \nu_n} \\
&\times {}_{\nu_1 \ldots \nu_n }\langle \dl_1+n | \cO(x_0)\ket{0} \bra{0} \widetilde{\cO}(x_0) \phi(x) q^{D} |\dl_1+n \rangle_{\mu_1\ldots \mu_n} \,,
\end{split}
\ee
where the trace $\Tr$ over the space of states $\cH$ is expanded into intermediate states of conformal dimensions $\dl_1$. The action of $\cO(x)$ on (translationally invariant) vacuum state $\ket{0}$ creates the sum over  descendant states: 
\be
\label{actonvac}
\begin{split}
\cO(x)\ket{0} &= e^{x \cdot P} \cO(0) e^{- x \cdot P} \ket{0} = \sum_{m=0}^{\infty} \frac{x^{\rho_1} \cdots x^{\rho_m}}{m!} P_{\rho_1} \cdots P_{\rho_m} \cO(0) \ket{0} \\
&= \sum_{m=0}^{\infty} \frac{x^{\rho_1} \cdots x^{\rho_m}}{m!} \ket{\dl+m}_{\rho_1 \ldots \rho_m}\,.
\end{split}
\ee
Then, it follows that
\begin{multline}
\left(B^{-1}_{\dl_1}\right)^{\mu_1 \ldots \mu_n ; \nu_1 \ldots \nu_n} {}_{\nu_1 \ldots \nu_n }\langle \dl_1+n | \cO(x_0)\ket{0} \\
=  \delta_{\dl \dl_1}  \frac{x_0^{\rho_1} \cdots x_0^{\rho_n}}{n!}\, (B^{-1}_{\dl_1})^{\mu_1 \ldots \mu_n ; \nu_1 \ldots \nu_n} \, (B_{\dl})_{\nu_1 \ldots \nu_n; \rho_1 \ldots \rho_n}
= \delta_{\dl \dl_1}  \frac{x_0^{\mu_1} \cdots x_0^{\mu_n}}{n!}\,,
\end{multline}
where we used that the descendant states  are orthogonal (by construction,  they are either at different  levels in the same module or  belong to different modules). Coming back to \eqref{trprojfirst} we see that
\be
\begin{split}
\Tr \left[ \Pi_{\dl} \phi(x) q^{D} \right] &= \int_{\mathbb{R}^D} {\rm d}^D x_0 \sum_{n=0}^{\infty}  \frac{x_0^{\mu_1} \cdots x_0^{\mu_n}}{n!} \, \bra{0} \widetilde{\cO}(x_0) \phi(x) q^{D} |\dl+n \rangle_{\mu_1\ldots \mu_n} \\
&= q^{\dl} \, \int_{\mathbb{R}^D} {\rm d}^D x_0 \sum_{n=0}^{\infty}  \frac{ q x_0^{\mu_1}\cdots q x_0^{\mu_n}}{n!} \, \bra{0} \widetilde{\cO}(x_0) \phi(x) |\dl+n  \rangle_{\mu_1\ldots \mu_n}\,.
\end{split}
\ee
Using \eqref{actonvac} we finally conclude that 
\be
\Tr \left[ \Pi_{\dl} \phi(x) q^{D} \right] = C_{\dl, h, \dl} \Upsilon_{\dl}^{h}(q,x)\,,
\ee
where we explicitly singled out a 1-point {\it thermal conformal partial wave} defined as 
\be
\label{tcpw1}
\Upsilon_{\dl}^{h} (q, x) =  N_{\dl} K_{\dl}^{h,\dl} \, q^{\dl}\, \int_{\mathbb{R}^D} {\rm d}^D x_0 V_{\td,h,\dl}(x_0, x, q x_0) \,.
\ee
When deriving this formula we used the relation between the structure constants \eqref{structconstratio}. Substituting the 3-point function \eqref{vert3} into \eqref{tcpw1} one obtains 
\be
\label{tcpw12}
\Upsilon_{\dl}^{h} (q, x) =  N_{\dl}\, K_{\dl}^{h,\dl}\, \frac{q^{D-h-\dl}}{(1-q)^{D-h}}\; T_{2}^{a_1,a_2; a_0}(x_1, x_2) \,,
\ee
where $x_1 = x/q$, $x_2 = x$, the parameters $a_i$ are expressed in terms of conformal dimensions:  
\be
\label{t2params}
a_1 = \frac{h+2\dl-D}{2}\,, \quad a_2 = \frac{D+h-2\dl}{2}\,, \quad a_0 = \frac{D-h}{2}\,.
\ee
Here, we defined a {\it thermal conformal integral}:
\be
\label{t2}
T_2^{a_1,a_2;a_0}(x_1,x_2) = \int_{\mathbb{R}^D} {\rm d}^D x_0 X_{01}^{-a_1}\, X_{02}^{-a_2}\, (x_0^2)^{-a_0}\,, \qquad a_1+a_2+2a_0 = D \,.
\ee
The constraint on parameters $a_i$ is satisfied due to  \eqref{t2params}.

\subsection{Thermal conformal integral}

It turns out that the thermal conformal integral \eqref{t2} is a particular case of the integrals explicitly calculated  by Boos and Davydychev in the context of one-loop Feynman integrals (with unconstrained parameters $a_i$) \cite{Boos:1987bg, Boos:1990rg}. Let us briefly describe their approach using the example \eqref{t2} and then suggest a different view of the same problem. 

\paragraph{One-loop Feynman integral.} Following \cite{Boos:1987bg, Boos:1990rg} we use the Schwinger parametrization 
\be
\label{schwingpar}
\frac{1}{X_{0i}^{a_i}} = \frac{1}{\Gamma(a_i)} \int_{0}^{\infty} \frac{{\rm d} \lambda_i}{\lambda_i} \lambda_i^{a_i} \operatorname{e}^{-\lambda_i X_{0i}} \,,
\ee
in order to replace the $D$-dimensional integral by a set of one-dimensional integrals:
\be
T_2^{a_1,a_2;a_0}(x_1,x_2) = \frac{\pi^{\hd}}{\Gamma(a_0,a_1,a_2)}\, \int_0^{\infty} \prod_{i=0}^2 \left( {\rm d}\lambda_i \lambda_i^{a_i-1}\right) \frac{1}{\Lambda^{\hd}} \, \operatorname{e}^{-\frac{1}{\Lambda}\left(\lambda_0\lambda_1 x_1^2 + \lambda_0 \lambda_2 x_2^2 + \lambda_1 \lambda_2 X_{12}\right)}\,,
\ee
where $\Lambda = \lambda_0 + \lambda_1 + \lambda_2$. Rescaling $\lambda_i \to \Lambda \lambda_i$ one  removes $1/\Lambda$ from the exponent and obtains  
\be
T_2^{a_1,a_2;a_0}(x_1,x_2) = \frac{2 \pi^{\hd}}{\Gamma(a_0,a_1,a_2)}\, \int_0^{\infty}   \prod_{i=0}^2 \left( {\rm d}\lambda_i \lambda_i^{a_i-1}\right) \frac{1}{\Lambda^{a_0}} \, \operatorname{e}^{-\lambda_0\lambda_1 x_1^2 -  \lambda_0 \lambda_2 x_2^2 - \lambda_1 \lambda_2 X_{12}}\,,
\ee
where $a_1+a_2+2a_0 = D$. Changing the integration variables as $\lambda_0 \to \sqrt{\lambda_0}$, $\lambda_1 \to \lambda_1 \sqrt{\lambda_0}$, $\lambda_2 \to \lambda_2 \sqrt{\lambda_0}$ and integrating over $\lambda_0$ one finds 
\be
T_2^{a_1,a_2;a_0}(x_1,x_2) = \frac{ \pi^{\hd}\, \Gamma(a'_0) }{ \Gamma(a_0, a_1, a_2) }\, \int_0^{\infty} \prod_{i=1}^2 \left( {\rm d}\lambda_i \lambda_i^{a_i-1}\right)\, \frac{(1+\lambda_1+\lambda_2)^{-a_0}}{(\lambda_1 x_1^2 +\lambda_2 x_2^2 + \lambda_1 \lambda_2 X_{12})^{a'_0}}\;.
\ee
Now, changing the integration variables as $\lambda_1 = s \sigma$, $\lambda_2 = s(1-\sigma)$, where $s\in (0,\infty)$, $\sigma \in (0,1)$ and using the integral representation for the Gauss function (equation 2.12(5) in \cite{Bateman:100233}) one finds that
\be
\label{t2compar}
T_2^{a_1,a_2;a_0}(x_1,x_2) = \pi^{\hd} \Gamma
\left[
\begin{array}{l l}
a'_0,\, a'_0 \\
a_1, a_2, \hd
\end{array}
\right]   \left(x_2^2\right)^{-a'_0} \int_{0}^1   \frac{{\rm d}\sigma \, \sigma^{a_1-1} (1-\sigma)^{a_2 - 1}}{\left(  1-\sigma\left(1-{x_1^2}/{x_2^2}\right) \right)^{a'_0}} {}_2 F_1 \left[
\begin{array}{l l}
a'_0, a'_0 \\
\quad \hd
\end{array}\bigg| 
1-\xi(\sigma)
\right],
\ee
where 
\be
\xi(\sigma) = \frac{\sigma (1-\sigma) X_{12}/x_2^2}{ 1-\sigma\left(1-{x_1^2}/{x_2^2}\right)}\,.
\ee
It can be shown that the integral \eqref{t2} can be expressed as a linear combination of the fourth Appell functions \cite{Boos:1987bg, Boos:1990rg}, which are defined as the following double series \cite{Bateman:100233} 
\be
\label{appel4}
F_4 \left[
\begin{array}{l l}
A_1, A_2  \\
 C_1, C_2
\end{array}\bigg| 
\xi_1, \xi_2
\right] 
= 
\sum_{m_1, m_2 = 0}^{\infty} \frac{(A_1)_{m_1+m_2} (A_2)_{m_2+m_1}  }{(C_1)_{m_1} (C_2)_{m_2} } \frac{\xi_1^{m_1} }{m_1!} \frac{\xi_2^{m_2} }{m_2!} \,,
\ee
converging when $\sqrt{|\xi_1|}+\sqrt{|\xi_2|} < 1$. 

\paragraph{Conformal integral.} However, it is more instructive to note that the thermal conformal integral \eqref{t2} can be obtained as a particular limit of the 4-point conformal integral:
\be
\label{i4}
I_4^{{\bm a}}({\bm x}) = \int_{\mathbb{R}^D} {\rm d}^D x_0 \prod_{i=1}^4 X_{0i}^{-a_i}\,, \qquad \sum_{j=1}^4 a_j =D \,,
\ee
where ${\bm a} = \{a_1,a_2,a_3,a_4 \}$ and ${\bm x} = \{x_1,x_2,x_3,x_4 \}$.\footnote{For non-parametric conformal integral (i.e. when all $a_i=1$ and, hence, $D=4$) the analogous relation was proved in \cite{Usyukina:1992jd} (see also \cite{Drummond:2006rz} for a related discussion). A similar relation for general $a_i$ is mentioned in \cite{Dolan:2000uw} (see footnote 9 therein).}  The constraint on parameters $a_i$ ensures that $I_4^{{\bm a}}({\bm x})$ is covariant under $O(D+1,1)$ conformal transformations. The $I_4^{{\bm a}}({\bm x})$ can be expressed in terms of special functions\footnote{This  conformal integral has been calculated by many authors, see e.g. Appendix of \cite{Dolan:2000uw} and references therein.} 
\be
\label{i4sum}
\begin{split}
I_4^{{\bm a}}({\bm x}) &= \frac{\pi^{\hd} \, \cL_4^{{\bm a}}({\bm x})} {\Gamma(a_1,a_2,a_3,a_4)}\, \bigg(v^{a'_4}\, J_4^{\{a_1,a_2,a_3,a_4\}}(u,v) + v^{a_1}\, J_4^{\{a_2,a_1,a_4,a_4\}}(u,v) \\
&+ u^{a_3+a_4-\hd} v^{a'_4} \, J_4^{\{a_4,a_3,a_2,a_1\}}(u,v) + u^{a_3+a_4-\hd} v^{a_1}\, J_4^{\{a_3,a_4,a_1,a_2\}}(u,v) \bigg) \,,
\end{split}
\ee
where the cross-ratios $u$ and $v$ are defined below in \eqref{cross4}, and\footnote{{When all $a_i=1$ and $D=4$ the 4-point conformal integral is expressed in terms of polylogarithms, see appendix \bref{app}.}
}
\be
\begin{split}
\label{i41appell}
J_4^{\{a_1,a_2,a_3,a_4\}}(u,v) &= \Gamma \left(a_2,a'_4,a_3-a'_4, a_1-a'_4 \right) \,
F_4 \left[
\begin{array}{l l}
\qquad a_2\,, \qquad a'_4 \\
1+a'_4-a_3\,, 1+a'_4-a_1
\end{array}\bigg|\, u\,, v\,
\right]\,.
\end{split}
\ee
The function $\cL_4^{{\bm a}}({\bm x})$ in \eqref{i4sum} is the leg-factor  responsible for the conformal covariance of the conformal integral:
\be
\label{leg4}
\cL_4^{{\bm a}}({\bm x}) = X_{14}^{-a_1} X_{24}^{\hd-a_2-a_4} X_{34}^{\hd-a_3-a_4} X_{23}^{a_4-\hd}\,.
\ee

By analogy with the previous paragraph,  one can show that the conformal integral \eqref{i4} can be represented as 
\be
\label{i4compar}
I_4^{{\bm a}}({\bm x}) = \pi^{\hd} \Gamma
\left[
\begin{array}{l l}
\, \, a'_3, a'_4\\
a_1, a_2, \hd
\end{array}
\right] \cL_4^{{\bm a}}({\bm x}) \int_{0}^{1} {\rm d}\sigma \frac{\sigma^{a_1-1} (1-\sigma)^{a_2-1}}{\left(1-\sigma \left( 1- 1/v\right) \right)^{a'_4}} \; {}_2 F_1 
\left[
\begin{array}{l l}
a'_3, a'_4 \\
\, \, \hd
\end{array}\bigg|\, 1-\xi(\sigma) \,
\right],
\ee
where $a_i' = D/2 - a_i$, the Gauss function's argument  and the cross-ratios are given by
\be
\label{cross4}
\xi(\sigma) = \frac{\sigma (1-\sigma) u/v}{1-\sigma \left( 1- 1/v\right)}\,, \qquad u = \frac{X_{12} X_{34}}{X_{13} X_{24}}\,, \qquad v = \frac{X_{14} X_{23}}{X_{13} X_{24}}\,.
\ee

Comparing \eqref{i4compar} and \eqref{t2compar} one  concludes that these two integrals are related to each other by taking a limit
\be
\label{i4t2conn}
T_2^{a_1,a_2;a_0}(x_1,x_2) = \lim_{\substack{x_3 \to 0 \\ \;\;x_4 \to \infty}} \left(X_{14}^{-a_1} X_{24}^{\hd-a_2-a_4} X_{34}^{\hd-a_3-a_4} \right)^{-1} I_4^{a_1,a_2,a_0,a_0}(x_1,x_2,x_3,x_4)\,.
\ee 
The limit corresponds to partial breaking conformal invariance of $I_4^{a_1,a_2,a_0,a_0}(x_1,x_2,x_3,x_4)$ by fixing two points $x_3 = 0$ and $x_4=\infty$. The residual symmetry of $T_2^{a_1,a_2;a_0}(x_1,x_2)$ is $O(1,1)\oplus O(D)$ which is the broken conformal  symmetry of the thermal correlation function  upon introducing one chemical potential (Hamiltonian/temperature), see \eqref{ward}.

\subsection{One-point thermal conformal block}
\label{sec:1pt_thermal}

Having   \eqref{i4sum} and \eqref{i4t2conn} we conclude that the thermal conformal partial wave \eqref{tcpw12} can be represented as a linear combination of  thermal conformal and shadow blocks 
\be
\label{tcpw1sum}
\Upsilon_{\dl}^{h}(q,x) = \cF_{\dl}^{h} (q,x) + K_{\dl}^{h,\dl} K_{\dl}^{\tilde{h},\dl} N_{\dl} \, \cF_{\td}^{h}(q,x) \,.
\ee
The 1-point thermal conformal block here is given by the product of two functions 
\be
\label{tblock}
\cF_{\dl}^{h}(q,x) = \cL^{h}(x) \cV_{\dl}^{h}(q)\,,
\ee
where the thermal leg factor $\cL^{h}(x) = r^{-h}$ 
solves the Ward identities \eqref{ward}. The bare conformal block is given by the fourth Appell function\footnote{The shadow block $\cF_{\td}^{h}$ is obtained from \eqref{tblock}-\eqref{tbare} by substituting $\dl \to \td$.}
\be
\label{tbare}
\ba{c}
\dps
\cV_{\dl}^{h}(q) =  \Gamma
\left[
\begin{array}{l l}
\dl, h-\hd \\
\frac{h}{2}, \dl - \frac{\widetilde{h}}{2}
\end{array}
\right]  \,q^{\dl}  (1-q)^{-h} \,  F_4 \bigg[
\begin{array}{l l}
\dl-\frac{h}{2}\,, \hd - \frac{h}{2} \\
1+\hd - h\,,  1-\hd +\dl
\end{array}\bigg| 
(1-q)^2\,, q^2
\bigg]
\vspace{3mm}
\\
\dps
\qquad\;\;+ (h \to \widetilde{h}) \,.
\ea
\ee
Note that the 1-point thermal block is invariant under permutation of $h$ and its shadow  $h \leftrightarrow \tilde{h}=D-h$: $\cV_{\dl}^{h}(q)=\cV_{\dl}^{\tilde{h}}(q)$.\footnote{In the case of CFT$_2$ such a  permutation symmetry can be seen from the Casimir differential equation \cite{Kraus:2017ezw}.}

\paragraph{Character.} Setting  $h=0$ in \eqref{tblock}-\eqref{tbare} yields the character of the scalar Verma module for the $D$-dimensional conformal algebra, 
\be
\label{chracter}
\cF_{\dl}^{h=0} = \frac{q^{\dl}}{(1-q)^D}\,,
\ee
which can be viewed as the $0$-point conformal block arising when expanding the thermal partition function (see e.g. \cite{Dolan:2005wy}).

\paragraph{CFT$_2$.} In two dimensions $D=2$ both Appell functions  in \eqref{tbare} factorize into the product of  Gauss functions \cite{Bateman:100233} 
\be
\ba{c}
\dps
\cV_{\dl}^{h}(q) = \frac{q^{\dl}}{(1-q)^h} \, {}_2 F_1 
\left[
\begin{array}{l l}
\dl - \frac{h}{2}, 1-\frac{h}{2} \\
\quad \dl
\end{array}\bigg| q
\right] \, \bigg(\Gamma
\left[
\begin{array}{l l}
\dl, h-1 \\
\frac{h}{2}, \dl + \frac{h}{2}-1
\end{array}
\right]  \, {}_2 F_1 
\left[
\begin{array}{l l}
\dl - \frac{h}{2}, 1-\frac{h}{2} \\
\quad 2-h
\end{array}\bigg| 1-q
\right] 
\vspace{3mm}
\\
\dps
\hspace{16mm}
+ \Gamma
\left[
\begin{array}{l l}
\dl, 1-h \\
1-\frac{h}{2}, \dl - \frac{h}{2}
\end{array}
\right]  \, (1-q)^{h-1} {}_2 F_1 
\left[
\begin{array}{l l}
\dl + \frac{h}{2}-1, \frac{h}{2} \\
\quad h
\end{array}\bigg| 1-q
\right]  \bigg)\,.
\ea
\ee
Then, for the expression in brackets, one can use the analytic continuation formula for the Gauss function \cite{Bateman:100233} (eq. 2.10(1) therein) to obtain the 1-point torus {\it global} conformal block \cite{Hadasz:2009db} 
\be
\label{tbare2dim}
\cV_{\dl}^{h}(q) = \frac{q^{\dl}}{(1-q)^h} \, {}_2 F_1 
\left[
\begin{array}{l l}
\dl - \frac{h}{2}, 1-\frac{h}{2} \\
\quad \dl
\end{array}\bigg| q
\right] \, {}_2 F_1 
\left[
\begin{array}{l l}
\dl - \frac{h}{2}, 1-\frac{h}{2} \\
\quad \dl
\end{array}\bigg| q
\right].
\ee

\paragraph{Low temperature.} 
Since $q = \exp(-\beta)$ then $q\to 0 $ corresponds to low temperatures $T \to 0$. In this regime the thermal block function behaves as  
\be
\label{asym_q}
\cV_{\dl}^{h}(q,x) \sim q^{\dl} \quad \mbox{  at }\, q \to 0 \,. 
\ee
It is this asymptotics which defines the splitting of the  four  terms in  \eqref{i4sum} into two groups of terms  in \eqref{tcpw1sum} relying on that the conformal and shadow blocks have different asymptotics at low temperatures. Indeed, the shadow block has the  asymptotics   \eqref{asym_q} but with the dual conformal dimension $\tilde\dl$ instead of $\Delta$. 

\paragraph{High temperature.}
In general, in the high-temperature regime $T \to \infty$ (i.e. $\beta \to 0$ or $q \to 1$) the thermal conformal block has two competing branches     
\be
\label{tblocklimit1}
\ba{c}
\dps
\cV_{\dl}^{h}(q) =  \beta^{-h} \, \Gamma
\left[
\begin{array}{l l}
\dl, h-\hd \\
\frac{h}{2}, \dl - \frac{\widetilde{h}}{2}
\end{array}
\right]   \,  {}_2 F_1 \bigg[
\begin{array}{l l}
\dl-\frac{h}{2}\,, \hd - \frac{h}{2} \\
1-\hd +\dl
\end{array}\bigg| 
 q^2
\bigg] + (h \to \widetilde{h})  \quad \text{ at } \, q\to 1\,.
\ea
\ee
Here, one of two terms dominates  depending on a value of the external conformal dimension $h$. To explicitly determine the limit of Gauss hypergeometric functions at $q \to 1$ one  distinguishes between  three cases  \cite{Bateman:100233}:
\be
\label{2f1limit}
{}_2 F_1\left[
\begin{array}{l l}
A, B \\
C
\end{array}\bigg| \xi
\right]  \underset{\xi \to 1}{\to} 		    \begin{cases}
    \Gamma\left[
    \begin{array}{l l}
    C, C-A-B \\
	C-A, C-B   
    \end{array}
    \right], \quad \text{ if }  		\re(C-A-B) > 0  \,,
    \\
     \vspace{0mm}
    (1-\xi)^{C-A-B} \Gamma\left[
    \begin{array}{l l}
    C, A-B-C \\
	A, B  
    \end{array}
    \right], \quad \text{ if } \re(C-A-B) < 0 \,,
    \vspace{0mm} 
    \\
    \ln(1/(1-\xi)) \Gamma\left[
    \begin{array}{l l}
    A + B \\
    A,B
    \end{array}
    \right], \quad \text{ if } C=A+B\,.
  \end{cases}
\ee
Applying these relations to each of the two hypergeometric functions in \eqref{tblocklimit1} one firstly notes that when  $h=D-1$ or $h=1$ the thermal block asymptotics is determined by the last line of \eqref{2f1limit} (we omit the $\Gamma$-prefactors):
\be
\label{tblocklog}
\cV_{\dl}^{h}(q) \sim \beta^{1-D} \ln \beta  \;\; \text{ at } \; \beta \to 0\,, \,\text{ when }\, h=D-1\, \text{ or }\, h=1 
\ee
The appearance of the logarithmic terms is not occasional since the associated hypergeometric-type functions at particular parameters can be expressed in terms of elementary functions and  polylogarithms. E.g. the non-parametric conformal and thermal integrals are given by the Bloch-Wigner function (see appendix \bref{app} for details and \cite{Dolan:2000ut} for a related discussion).   
     
Beyond these cases, there are three possible asymptotics of the thermal conformal block: 
\be
\label{tblocklimit2}
\cV_{\dl}^{h}(q) \sim \beta^{-\operatorname{max}(h,\,D-h,\,D-1)} \;\; \text{ at }\; \beta \to 0\,,
\ee
which result from combining  the first two lines in the list   \eqref{2f1limit} inside \eqref{tblocklimit1}. Note that the asymptotics $\beta^{1-D}$ is possible only for $D > 2$.\footnote{This analysis of the three different asymptotics agrees with the results previously discussed in \cite{Gobeil:2018fzy, Iliesiu:2018fao}.}

As an example consider $h>D-1\geq 1$, when the thermal conformal block asymptotics is given by
\be
\label{tblocklimit3}
\cV_{\dl}^{h>D-1\geq 1}(q) = \beta^{-h} \, \Gamma 
\left[
\begin{array}{l l}
\dl, 1+\dl-\hd, 1+h-D, h-\hd \\
\frac{h}{2}, 1-\frac{\widetilde{h}}{2}, 1+\dl+\frac{h}{2}-D, \dl-\frac{\widetilde{h}}{2}
\end{array}
\right]\,, \qquad \beta\to 0\,.
\ee
For $D=2$ this agrees with the two-dimensional case considered in \cite{Iliesiu:2018fao}. Having an explicit form of the high-temperature asymptotics  of the thermal conformal block, it is possible to write the contribution of scalar operators to the constant $b_h$, which defines the 1-point thermal correlation function on $S^1_{\beta} \times \mathbb{R}^{D-1}$ \eqref{bh} as
\be
b_h = \frac{1}{Z(0)} \sum_{\dl} C_{\dl,h,\dl}\Gamma 
\left[
\begin{array}{l l}
\dl, 1+\dl-\hd, 1+h-D, h-\hd \\
\frac{h}{2}, 1-\frac{\widetilde{h}}{2}, 1+\dl+\frac{h}{2}-D, \dl-\frac{\widetilde{h}}{2}
\end{array}
\right]\,.
\ee

\paragraph{Selfdual conformal dimensions.} When the internal conformal dimension equals its shadow $\dl = \td$ (i.e. $\dl = D/2$), the two conformal blocks in the decomposition  \eqref{tcpw1sum} coincide $\cF_{\dl}^h=\cF_{\td}^{h}$ and the thermal conformal partial wave is zero
\be
\Upsilon_{\dl=\td}^h(q,x) = 0\,, \quad \text{ since }  \quad K_{\dl}^{h,\dl} K_{\dl}^{\tilde{h},\dl} N_{\dl}\bigg|_{\dl=\td} = -1\,.
\ee
This happens due to the presence of $N_{\dl}$ prefactor in the definition of the thermal conformal partial wave \eqref{tcpw12}, which is equal to zero at $\dl=D/2$. 

Nevertheless, the thermal conformal block is not vanishing. For concreteness, let $D=4$ and $h=\dl=2$. This case is curious because the parameters \eqref{t2params} $a_i=1$ and the conformal integral \eqref{i4} as well as the corresponding thermal conformal integral \eqref{t2compar} are expressed through the polylogarithms (see appendix \bref{app}, and, in particular, eqs. \eqref{i41111} and \eqref{t2111}). The thermal conformal block is given by 
\be
\label{tblock224}
\cV_{\dl=2}^{h=2}(q)\Big|_{D=4} =\frac{q^2}{(1-q)^3} \,.
\ee
Its asymptotics at $\beta \to 0$  is $(1-q)^{-3} = \beta^{1-D}$, which demonstrates the general analysis \eqref{tblocklimit2} from the previous paragraph.

Finally, note that in Ref. \cite{Gobeil:2018fzy} the 1-point thermal conformal block was calculated by means of the conjectured AdS-integral representation. The result was given in terms of the generalized hypergeometric function ${}_3 F_2$ and it would be important to prove that the two results coincide. However, that will require knowledge of special identities between the generalized hypergeometric  ${}_3 F_2$ and the   Appell $F_4$ functions which are unknown to us. Nonetheless, using the Wolfram Mathematica  we managed to verify the coincidence of the two functions for particular values of parameters $\dl, h$, $D$ up to $O(q^4)$.

\section{Conclusion}
\label{sec:concl}

In this paper we have elaborated on the shadow formalism in thermal CFT$_D$ on $\cylS$.  By way of illustration, we have explicitly calculated the 1-point thermal conformal block for scalar operators which proved to be the fourth Appell function.  

The new ingredient introduced in this paper is the thermal conformal integral which defines the corresponding  thermal conformal partial wave. It differs from the conventional conformal integrals by having a different constraint on parameters. It is known that, being a linear combination of the conformal and shadow blocks, the 4-point conformal partial wave is proportional to the 4-point conformal integral \eqref{i4} \cite{SimmonsDuffin:2012uy, Rosenhaus:2018zqn}. On the other hand, by means of the limit relation \eqref{i4t2conn} the same conformal integral defines the sum of  thermal and shadow conformal blocks. Thus, we conclude that the 4-point conformal integral may be considered as the basic object which particular limits produces either 4-point conformal blocks on $\mathbb{R}^D$ or 1-point thermal blocks on $\mathbb{S}^1_{\beta} \times \mathbb{S}^{D-1}$.\footnote{It is worth noting that conformal integrals  in the context of thermal \cft were  recently considered in Ref.  \cite{Petkou:2021zhg}, where they were shown to define 1-point thermal correlation functions  of certain operators on $S^1_{\beta} \times R^{D-1}$ (see also related works \cite{Petkou:2018ynm,Karydas:2023ufs}). The conformal integrals were also proved to be  Yangian invariant \cite{Chicherin:2017cns, Chicherin:2017frs}.} 

It would be important to understand which $o(D+1,1)$ Casimir equation is satisfied by the thermal 1-point block. In two dimensions this problem was solved for torus blocks in any  channel  \cite{Alkalaev:2022kal,Alkalaev:2023evp,Kraus:2017ezw}. In $D$ dimensions the (plane) conformal partial waves obey the Casimir equations by construction \cite{SimmonsDuffin:2012uy}. It would be interesting to check this property for thermal partial waves introduced in this paper. On the other hand, the Casimir equation for correlation functions with more than one chemical potentials was proposed in Ref. \cite{Gobeil:2018fzy}. Nonetheless, the Casimir equation for the 1-point thermal  correlation function \eqref{corrcyl} is still unknown.

Finally, let us note that the present thermal shadow formalism can be naturally generalized  to include spinning operators as well as to calculate $n$-point  thermal scalar blocks \cite{AM}. Both of these directions are potentially   important  for  understanding conformal bootstrap at finite temperature.

\vspace{3mm}

\noindent \textbf{Acknowledgements.} We are grateful to Vladimir Khiteev  for useful discussions and Anastasios Petkou for interesting correspondence. S.M. also thanks Vagif Tagiev for discussions on conformal integrals.  Our work was supported by the Foundation for the Advancement of Theoretical Physics and Mathematics “BASIS”.

\appendix

\section{Non-parametric conformal integral}
\label{app}

Consider the 4-point {\it non-parametric} conformal integral \eqref{i4sum} which is defined by choosing unit parameters $a_i=1$ in $D=4$. The individual term \eqref{i41appell} diverges  due to the poles of the $\Gamma$-functions prefactor. In order to regularize the singularity one introduces a regularization parameter $\epsilon \to 0$ as follows 
\be
\label{refparam}
a_1=a_2=a_3=1\,, \qquad a_4 = 1-2\epsilon \qquad \Rightarrow \qquad D=4-2\epsilon\,.
\ee
Applying this regularization inside \eqref{i4sum} and expanding it around  $\epsilon = 0$ one can verify that
\be
I_4^{{\bm 1}_{\epsilon}}({\bm x}) = \frac{\pi^2}{X_{13} X_{24}} \frac{ \Phi(u,v)}{\lambda(u,v)} + O(\epsilon)\,,
\ee
where ${\bm 1}_{\epsilon} = \{1,1,1,1-2\epsilon \}$, $\lambda(u,v) = \sqrt{(1-u-v)^2-4 u v}$ and the Bloch-Wigner function $\Phi(u,v)$ is expressed in terms of  polylogarithms \cite{Davydychev:1992eww}: 
\be
\label{blochuv}
\begin{split}
\Phi(u,v) &=  \frac{\pi^2}{3} + \ln u \ln v + \ln\left(\frac{1+u-v-\lambda(u,v)}{2u} \right) \ln\left(\frac{1-u+v-\lambda(u,v)}{2v} \right)   \\
&+ 2\ln\left(\frac{1+u-v-\lambda(u,v)}{2u} \right) + 2 \ln\left(\frac{1-u+v-\lambda(u,v)}{2v} \right) \\
&- 2\Li2\left(\frac{1+u-v-\lambda(u,v)}{2} \right) - 2\Li2\left(\frac{1-u+v-\lambda(u,v)}{2} \right) \,,
\end{split}
\ee
where $\Li2$ is a dilogarithm. It is convenient to introduce new variables $z$ and $y$:
\be
\label{crosszy}
u = z(1-y) \,, \qquad v = y(1-z)\,,
\ee
in terms of which one has $\lambda(z,y) = 1-y-z$ and the Bloch-Wigner function \eqref{blochuv} takes the form (see \cite{Dolan:2000uw})
\be
\label{blochzy}
\Phi(z,y) =  \ln(y(1-z)) \ln\left(\frac{z}{1-y}\right) + 2 \Li2(1-z) -2 \Li2 y \,,
\ee
where we have used the identity for the dilogarithm $\Li2 z + \Li2(1-z) = \pi^2/6-\ln z \ln(1-z)$. This function allows rewriting the 4-point non-parametric conformal integral as
\be
\label{i41111}
I_4^{1,1,1,1}({\bm x}) = \frac{\pi^2}{X_{13} X_{24}} \frac{ \Phi(z,y)}{\lambda(z,y)}\,,
\ee
where $z$ and $y$ depend on  ${\bm x}$. 

Let us now consider a subspace in the $(y,z)$-space \eqref{crosszy} singled out  by $\lambda(z,y) = 0$. The later constraint defines a continuous line of  poles of the non-parametric conformal integral \eqref{i41111}. Nonetheless, one can examine the form of $I_4^{1,1,1,1}({\bm x})$ near the singularity by introducing  a regularization parameter $\alpha$ as follows 
\be
\label{yreg}
y_{\alpha} = \alpha(1-z)\qquad \Rightarrow \qquad \lambda(z,y_{\alpha}) = (1-z)(1-\alpha)\,.
\ee
Expanding $\Phi(z,y)/\lambda(z,y)$ near $\alpha=1$ yields
\be
\label{blochsubspace}
\frac{\Phi(z,y_{\alpha})}{\lambda(z,y_{\alpha})} = -\frac{2\ln(z)}{1-z}-\frac{2\ln(1-z)}{z} + O(\alpha-1)\,,
\ee
which in turn defines the corresponding conformal integral \eqref{i41111} on the subspace $\lambda(z,y)=0$. 

The constraint $\lambda(z,y) = 0$ is relevant when considering the thermal conformal blocks in which case $u=q^2$ and $v = (1-q)^2$ (see \eqref{tbare}). Using the relation between the conformal and thermal  integrals \eqref{i4t2conn}  one finds that 
\be
\label{t2111}
T_2^{1,1;1}(x/q,x) = -2\pi^2 q^2 r^{-2} \left(\frac{\ln q}{1-q}+\frac{\ln(1-q)}{q}\right)\,,
\ee

In terms of conformal dimensions the choice $a_i=1$ corresponds to $h = \Delta =2$ and $D=4$ (see \eqref{t2params}). In order to examine the corresponding thermal conformal block one  introduces first a function of two variables  
\be
\label{tbareuv}
\ba{c}
\dps
\cV_{\dl}^{h}(u,v) =  \Gamma
\left[
\begin{array}{l l}
\dl, h-\hd \\
\frac{h}{2}, \dl - \frac{\widetilde{h}}{2}
\end{array}
\right]  \,u^{\dl/2}  v^{-h/2} \,  F_4 \bigg[
\begin{array}{l l}
\dl-\frac{h}{2}\,, \hd - \frac{h}{2} \\
1-\hd +\dl\,, 1+\hd - h
\end{array}\bigg| 
u\,, v
\bigg]
\vspace{3mm}
\\
\dps
\qquad\;\;+ (h \to \widetilde{h}) \,,
\ea
\ee
which reduces to the thermal conformal block \eqref{tbare} at $u=q^2$ and $v = (1-q)^2$. However, the $\Gamma$-prefactors here diverge at $h = \Delta =2$ and $D=4$, so to treat this expression one  again introduces a regularization parameter $\eta \to 0$ as $D=4-2\eta$. Expanding  \eqref{tbareuv} around $\eta=0$ one obtains  
\be
\cV_{\dl=2}^{h=2}(z,y)\bigg|_{D=4-2\eta} =- \frac{1}{1-z-y} \,\frac{z(1-y)}{y(1-z)}\,  \ln \frac{y}{1-z} + O(\eta)\,,
\ee
where $z$ and $y$ are related to   $u$ and $v$ as in \eqref{crosszy}. To consider the case $u=q^2$ and $v = (1-q)^2$ ($z=q$ and $y=1-q$) one uses a regularization as in \eqref{yreg}.  Expanding the last equation around $\alpha = 1$ one finds the 1-point thermal conformal block for $h=\dl=2$ in $D=4$  given in the equation \eqref{tblock224}.


\begin{thebibliography}{10}

\bibitem{Witten:1998zw}
E.~Witten, \emph{{Anti-de Sitter space, thermal phase transition, and
  confinement in gauge theories}},
  \href{http://dx.doi.org/10.4310/ATMP.1998.v2.n3.a3}{\emph{Adv. Theor. Math.
  Phys.} {\bf 2} (1998) 505--532},
  [\href{http://arxiv.org/abs/hep-th/9803131}{{\tt hep-th/9803131}}].

\bibitem{Poland:2018epd}
D.~Poland, S.~Rychkov and A.~Vichi, \emph{{The Conformal Bootstrap: Theory,
  Numerical Techniques, and Applications}},
  \href{http://dx.doi.org/10.1103/RevModPhys.91.015002}{\emph{Rev. Mod. Phys.}
  {\bf 91} (2019) 015002}, [\href{http://arxiv.org/abs/1805.04405}{{\tt
  1805.04405}}].

\bibitem{Bissi:2022mrs}
A.~Bissi, A.~Sinha and X.~Zhou, \emph{{Selected Topics in Analytic Conformal
  Bootstrap: A Guided Journey}},  \href{http://arxiv.org/abs/2202.08475}{{\tt
  2202.08475}}.

\bibitem{Dolan:2011dv}
F.~Dolan and H.~Osborn, \emph{{Conformal Partial Waves: Further Mathematical
  Results}},  \href{http://arxiv.org/abs/1108.6194}{{\tt 1108.6194}}.

\bibitem{Costa:2011dw}
M.~S. Costa, J.~Penedones, D.~Poland and S.~Rychkov, \emph{{Spinning Conformal
  Blocks}}, \href{http://dx.doi.org/10.1007/JHEP11(2011)154}{\emph{JHEP} {\bf
  11} (2011) 154}, [\href{http://arxiv.org/abs/1109.6321}{{\tt 1109.6321}}].

\bibitem{Karateev:2017jgd}
D.~Karateev, P.~Kravchuk and D.~Simmons-Duffin, \emph{{Weight Shifting
  Operators and Conformal Blocks}},
  \href{http://dx.doi.org/10.1007/JHEP02(2018)081}{\emph{JHEP} {\bf 02} (2018)
  081}, [\href{http://arxiv.org/abs/1706.07813}{{\tt 1706.07813}}].

\bibitem{Rosenhaus:2018zqn}
V.~Rosenhaus, \emph{{Multipoint Conformal Blocks in the Comb Channel}},
  \href{http://dx.doi.org/10.1007/JHEP02(2019)142}{\emph{JHEP} {\bf 02} (2019)
  142}, [\href{http://arxiv.org/abs/1810.03244}{{\tt 1810.03244}}].

\bibitem{Fortin:2019zkm}
J.-F. Fortin, W.~Ma and W.~Skiba, \emph{{Higher-Point Conformal Blocks in the
  Comb Channel}}, \href{http://dx.doi.org/10.1007/JHEP07(2020)213}{\emph{JHEP}
  {\bf 07} (2020) 213}, [\href{http://arxiv.org/abs/1911.11046}{{\tt
  1911.11046}}].

\bibitem{Poland:2021xjs}
D.~Poland and V.~Prilepina, \emph{{Recursion relations for 5-point conformal
  blocks}}, \href{http://dx.doi.org/10.1007/JHEP10(2021)160}{\emph{JHEP} {\bf
  10} (2021) 160}, [\href{http://arxiv.org/abs/2103.12092}{{\tt 2103.12092}}].

\bibitem{Buric:2021ywo}
I.~Buric, S.~Lacroix, J.~A. Mann, L.~Quintavalle and V.~Schomerus,
  \emph{{Gaudin models and multipoint conformal blocks: general theory}},
  \href{http://dx.doi.org/10.1007/JHEP10(2021)139}{\emph{JHEP} {\bf 10} (2021)
  139}, [\href{http://arxiv.org/abs/2105.00021}{{\tt 2105.00021}}].

\bibitem{Ammon:2024axd}
M.~Ammon, J.~Hollweck, T.~H\"ossel and K.~W\"olfl, \emph{{Conformal Blocks in
  Two and Four Dimensions from Oscillator Representations}},
  \href{http://arxiv.org/abs/2406.19436}{{\tt 2406.19436}}.

\bibitem{Iliesiu:2018fao}
L.~Iliesiu, M.~Kolo\u{g}lu, R.~Mahajan, E.~Perlmutter and D.~Simmons-Duffin,
  \emph{{The Conformal Bootstrap at Finite Temperature}},
  \href{http://dx.doi.org/10.1007/JHEP10(2018)070}{\emph{JHEP} {\bf 10} (2018)
  070}, [\href{http://arxiv.org/abs/1802.10266}{{\tt 1802.10266}}].

\bibitem{Gobeil:2018fzy}
Y.~Gobeil, A.~Maloney, G.~S. Ng and J.-q. Wu, \emph{{Thermal Conformal
  Blocks}}, \href{http://dx.doi.org/10.21468/SciPostPhys.7.2.015}{\emph{SciPost
  Phys.} {\bf 7} (2019) 015}, [\href{http://arxiv.org/abs/1802.10537}{{\tt
  1802.10537}}].

\bibitem{Manenti:2019wxs}
A.~Manenti, \emph{{Thermal CFTs in momentum space}},
  \href{http://dx.doi.org/10.1007/JHEP01(2020)009}{\emph{JHEP} {\bf 01} (2020)
  009}, [\href{http://arxiv.org/abs/1905.01355}{{\tt 1905.01355}}].

\bibitem{Ferrara:1972ay}
S.~Ferrara, A.~F. Grillo and G.~Parisi, \emph{{Nonequivalence between conformal
  covariant wilson expansion in euclidean and minkowski space}},
  \href{http://dx.doi.org/10.1007/BF02815915}{\emph{Lett. Nuovo Cim.} {\bf 5S2}
  (1972) 147--151}.

\bibitem{Ferrara:1972xe}
S.~Ferrara and G.~Parisi, \emph{{Conformal covariant correlation functions}},
  \href{http://dx.doi.org/10.1016/0550-3213(72)90480-4}{\emph{Nucl. Phys. B}
  {\bf 42} (1972) 281--290}.

\bibitem{Ferrara:1972uq}
S.~Ferrara, A.~F. Grillo, G.~Parisi and R.~Gatto, \emph{{The shadow operator
  formalism for conformal algebra. Vacuum expectation values and operator
  products}}, \href{http://dx.doi.org/10.1007/BF02907130}{\emph{Lett. Nuovo
  Cim.} {\bf 4S2} (1972) 115--120}.

\bibitem{Ferrara:1972kab}
S.~Ferrara, A.~F. Grillo, G.~Parisi and R.~Gatto, \emph{{Covariant expansion of
  the conformal four-point function}},
  \href{http://dx.doi.org/10.1016/0550-3213(73)90467-7}{\emph{Nucl. Phys. B}
  {\bf 49} (1972) 77--98}.

\bibitem{Alkalaev:2023evp}
K.~Alkalaev and S.~Mandrygin, \emph{{Torus shadow formalism and exact global
  conformal blocks}},
  \href{http://dx.doi.org/10.1007/JHEP11(2023)157}{\emph{JHEP} {\bf 11} (2023)
  157}, [\href{http://arxiv.org/abs/2307.12061}{{\tt 2307.12061}}].

\bibitem{Belavin:2024nnw}
V.~Belavin and J.~Ramos~Cabezas, \emph{{Global conformal blocks via shadow
  formalism}}, \href{http://dx.doi.org/10.1007/JHEP02(2024)167}{\emph{JHEP}
  {\bf 02} (2024) 167}, [\href{http://arxiv.org/abs/2401.02580}{{\tt
  2401.02580}}].

\bibitem{El-Showk:2011yvt}
S.~El-Showk and K.~Papadodimas, \emph{{Emergent Spacetime and Holographic
  CFTs}}, \href{http://dx.doi.org/10.1007/JHEP10(2012)106}{\emph{JHEP} {\bf 10}
  (2012) 106}, [\href{http://arxiv.org/abs/1101.4163}{{\tt 1101.4163}}].

\bibitem{Marchetto:2023fcw}
E.~Marchetto, A.~Miscioscia and E.~Pomoni, \emph{{Broken (super) conformal Ward
  identities at finite temperature}},
  \href{http://dx.doi.org/10.1007/JHEP12(2023)186}{\emph{JHEP} {\bf 12} (2023)
  186}, [\href{http://arxiv.org/abs/2306.12417}{{\tt 2306.12417}}].

\bibitem{Petkou:2021zhg}
A.~C. Petkou, \emph{{Thermal one-point functions and single-valued
  polylogarithms}},
  \href{http://dx.doi.org/10.1016/j.physletb.2021.136467}{\emph{Phys. Lett. B}
  {\bf 820} (2021) 136467}, [\href{http://arxiv.org/abs/2105.03530}{{\tt
  2105.03530}}].

\bibitem{Fradkin:1978pp}
E.~S. Fradkin and M.~Y. Palchik, \emph{{Recent Developments in Conformal
  Invariant Quantum Field Theory}},
  \href{http://dx.doi.org/10.1016/0370-1573(78)90172-2}{\emph{Phys. Rept.} {\bf
  44} (1978) 249--349}.

\bibitem{Fradkin:1996is}
E.~S. Fradkin and M.~Y. Palchik, \emph{{Conformal quantum field theory in
  D-dimensions}}.
\newblock 1996.

\bibitem{SimmonsDuffin:2012uy}
D.~Simmons-Duffin, \emph{{Projectors, Shadows, and Conformal Blocks}},
  \href{http://dx.doi.org/10.1007/JHEP04(2014)146}{\emph{JHEP} {\bf 04} (2014)
  146}, [\href{http://arxiv.org/abs/1204.3894}{{\tt 1204.3894}}].

\bibitem{Symanzik:1972wj}
K.~Symanzik, \emph{{On Calculations in conformal invariant field theories}},
  \href{http://dx.doi.org/10.1007/BF02824349}{\emph{Lett. Nuovo Cim.} {\bf 3}
  (1972) 734--738}.

\bibitem{Boos:1987bg}
E.~E. Boos and A.~I. Davydychev, \emph{{A Method of the Evaluation of the
  Vertex Type Feynman Integrals}}, {\emph{Moscow Univ. Phys. Bull.} {\bf 42N3}
  (1987) 6--10}.

\bibitem{Boos:1990rg}
E.~E. Boos and A.~I. Davydychev, \emph{{A Method of evaluating massive Feynman
  integrals}}, \href{http://dx.doi.org/10.1007/BF01016805}{\emph{Theor. Math.
  Phys.} {\bf 89} (1991) 1052--1063}.

\bibitem{Bateman:100233}
H.~Bateman and A.~Erdélyi, \emph{{Higher transcendental functions}}.
\newblock California Institute of technology. Bateman Manuscript project.
  McGraw-Hill, New York, NY, 1953.

\bibitem{Usyukina:1992jd}
N.~I. Usyukina and A.~I. Davydychev, \emph{{An Approach to the evaluation of
  three and four point ladder diagrams}},
  \href{http://dx.doi.org/10.1016/0370-2693(93)91834-A}{\emph{Phys. Lett. B}
  {\bf 298} (1993) 363--370}.

\bibitem{Drummond:2006rz}
J.~M. Drummond, J.~Henn, V.~A. Smirnov and E.~Sokatchev, \emph{{Magic
  identities for conformal four-point integrals}},
  \href{http://dx.doi.org/10.1088/1126-6708/2007/01/064}{\emph{JHEP} {\bf 01}
  (2007) 064}, [\href{http://arxiv.org/abs/hep-th/0607160}{{\tt
  hep-th/0607160}}].

\bibitem{Dolan:2000uw}
F.~A. Dolan and H.~Osborn, \emph{{Implications of N=1 superconformal symmetry
  for chiral fields}},
  \href{http://dx.doi.org/10.1016/S0550-3213(00)00553-8}{\emph{Nucl. Phys. B}
  {\bf 593} (2001) 599--633}, [\href{http://arxiv.org/abs/hep-th/0006098}{{\tt
  hep-th/0006098}}].

\bibitem{Kraus:2017ezw}
P.~Kraus, A.~Maloney, H.~Maxfield, G.~S. Ng and J.-q. Wu, \emph{{Witten
  Diagrams for Torus Conformal Blocks}},
  \href{http://dx.doi.org/10.1007/JHEP09(2017)149}{\emph{JHEP} {\bf 09} (2017)
  149}, [\href{http://arxiv.org/abs/1706.00047}{{\tt 1706.00047}}].

\bibitem{Dolan:2005wy}
F.~A. Dolan, \emph{{Character formulae and partition functions in higher
  dimensional conformal field theory}},
  \href{http://dx.doi.org/10.1063/1.2196241}{\emph{J. Math. Phys.} {\bf 47}
  (2006) 062303}, [\href{http://arxiv.org/abs/hep-th/0508031}{{\tt
  hep-th/0508031}}].

\bibitem{Hadasz:2009db}
L.~Hadasz, Z.~Jaskolski and P.~Suchanek, \emph{{Recursive representation of the
  torus 1-point conformal block}},
  \href{http://dx.doi.org/10.1007/JHEP01(2010)063}{\emph{JHEP} {\bf 01} (2010)
  063}, [\href{http://arxiv.org/abs/0911.2353}{{\tt 0911.2353}}].

\bibitem{Dolan:2000ut}
F.~A. Dolan and H.~Osborn, \emph{{Conformal four point functions and the
  operator product expansion}},
  \href{http://dx.doi.org/10.1016/S0550-3213(01)00013-X}{\emph{Nucl. Phys. B}
  {\bf 599} (2001) 459--496}, [\href{http://arxiv.org/abs/hep-th/0011040}{{\tt
  hep-th/0011040}}].

\bibitem{Petkou:2018ynm}
A.~C. Petkou and A.~Stergiou, \emph{{Dynamics of Finite-Temperature Conformal
  Field Theories from Operator Product Expansion Inversion Formulas}},
  \href{http://dx.doi.org/10.1103/PhysRevLett.121.071602}{\emph{Phys. Rev.
  Lett.} {\bf 121} (2018) 071602}, [\href{http://arxiv.org/abs/1806.02340}{{\tt
  1806.02340}}].

\bibitem{Karydas:2023ufs}
M.~Karydas, S.~Li, A.~C. Petkou and M.~Vilatte, \emph{{Conformal Graphs as
  Twisted Partition Functions}},
  \href{http://dx.doi.org/10.1103/PhysRevLett.132.231601}{\emph{Phys. Rev.
  Lett.} {\bf 132} (2024) 231601}, [\href{http://arxiv.org/abs/2312.00135}{{\tt
  2312.00135}}].

\bibitem{Chicherin:2017cns}
D.~Chicherin, V.~Kazakov, F.~Loebbert, D.~M\"uller and D.-l. Zhong,
  \emph{{Yangian Symmetry for Bi-Scalar Loop Amplitudes}},
  \href{http://dx.doi.org/10.1007/JHEP05(2018)003}{\emph{JHEP} {\bf 05} (2018)
  003}, [\href{http://arxiv.org/abs/1704.01967}{{\tt 1704.01967}}].

\bibitem{Chicherin:2017frs}
D.~Chicherin, V.~Kazakov, F.~Loebbert, D.~M\"uller and D.-l. Zhong,
  \emph{{Yangian Symmetry for Fishnet Feynman Graphs}},
  \href{http://dx.doi.org/10.1103/PhysRevD.96.121901}{\emph{Phys. Rev. D} {\bf
  96} (2017) 121901}, [\href{http://arxiv.org/abs/1708.00007}{{\tt
  1708.00007}}].

\bibitem{Alkalaev:2022kal}
K.~Alkalaev, S.~Mandrygin and M.~Pavlov, \emph{{Torus conformal blocks and
  Casimir equations in the necklace channel}},
  \href{http://dx.doi.org/10.1007/JHEP10(2022)091}{\emph{JHEP} {\bf 10} (2022)
  091}, [\href{http://arxiv.org/abs/2205.05038}{{\tt 2205.05038}}].

\bibitem{AM}
K.~Alkalaev and S.~Mandrygin, \emph{{to appear}}.

\bibitem{Davydychev:1992eww}
A.~I. Davydychev, \emph{{Recursive algorithm for evaluating vertex-type Feynman
  integrals}}, \href{http://dx.doi.org/10.1088/0305-4470/25/21/017}{\emph{J.
  Phys. A} {\bf 25} (1992) 5587}.

\end{thebibliography}

\providecommand{\href}[2]{#2}\begingroup\raggedright\endgroup

\end{document}